\documentclass[a4]{article}
\usepackage{graphicx}
\newcommand{\be}{\begin{equation}}
\newcommand{\ee}{\end{equation}}
\renewcommand{\k}[2]{\frac{#1}{#2}}

\newcommand{\pcd}[2]{\frac{d^n \! #1}{d #2^n}}

\def\s{\,\,\,\,}
\def\t{\tau}
\def\d{\delta}
\def\q{\Gamma_{d}}
\def\z{\zeta}
\def\ra{\rightarrow}
\def\sumj{\sum_{j=1}^{\t/\d}}
\def\sumjj{\sum_{j=1}^{\infty}}
\def\sumr{\sum_{j=1}^{A r/(\xi \d)}}

\begin{document}
\title{\Large \bf Small-World Networks in Geophysics}

\author{  Xin-She Yang \\ 
{\small  Department of Applied Mathematics},\\
  {\small      University of Leeds, LEEDS LS2 9JT, UK}}

\date{}
\maketitle

\abstract{
Many geophysical processes can be modelled by using interconnected
networks. The small-world network model has recently attracted much attention
in physics and applied sciences. In this paper, we try to use and modify
the small-world theory to model geophysical processes such as
diffusion and transport in disordered porous rocks. We develop an analytical
approach as well as numerical simulations to try to characterize
the pollutant transport and percolation properties
of small-world networks. The analytical expression of system
saturation time and fractal dimension of small-world networks
are given and thus compared with numerical simulations. }


\noindent {\bf Citation Detail}: 
X. S. Yang, Small-world networks in geophysics, 
{\it Geophysical Research Letters}, {\bf 28} (13), 2549-2552 (2001).

\section{Introduction}

Many geophysical processes such as flow in disordered porous media,
magma transport and saltwater intrusion [{\it Holzbercher},1998]
can be modelled as percolation and diffusion
in interconnected networks that can generally be represented
as connected graphs. The properties of complicated networks
are mainly determined by the way of connections between occupied sites
or vertices. Two commonly used quantities, which characterise a graph, are
the degree of local clustering and the average distance between
vertices. The generic structure and the degree of the randomness
of networks have strong influence on the overal properties.
Obviously, regular network is the one limiting case with
a high degree of local clustering and a very large average distance,
while the other limiting case is the random network which shows negligible
local clustering and the average distance is relative small.
In reality, most networks are neither exactly regular nor completely random,
realistic networks are something between these two extremes.

Small-world network is a special kind of networks with a higher
degree of clustering and a small average distance. The degree of
the randomness of networks often has a strong influence on the
network properties. Recently, {\it Watts and Strogatz} [1998]
has presented a mathematical model to characterises the network structure
and to study the dynamic properties of small-world networks.
Such small-world phenomenon can be obtained by adding randomly only
a small fraction  of the connections, and some common networks such
as power grids, film stars networks and neural networks behaves like
small-world networks [{\it Newman and Watts}, 1999; {\it Moukarzel},1999;
{\it Newmann et al.}, 2000, {\it Boots and Sasaki}, 1999;
{\it Barthelemy and Amaral}, 1999].
Most recent studies focus on the spreading and shortest paths
in system with sparse long-range connections by using the small-world model
[{\it Newman and Watts}, 1999; {\it Moukarzel},1999; {\it Pandit
and Amritkar}, 1999].
The spreading of some influence such as a forest fire, an infectious disease
or a particle in percolating media is studied using a simple rule: at each
time step, the influence propagates from the infected site to all uninfected
sites connected to it via a link although this link is not necessary physical.
A long-range connection or shortcut can simulate the spark that starts a new
fire spot, the infect site (say person with flu) suddenly travel to a new
place or site, or a portable computer with virus that start to connect to the
network a new place. These research mainly focus on
the immediate response of the network by adding randomly some
long-distance shortcuts, there is no time delay in the
network systems. However, in reality, a spark or an
infection can not start a new fire spot or new infection immediately,
it usually takes some time $\Delta$, called ignition time or waiting time,
to start a new fire or infection. Thus the existing models are
no longer be able  to predict the response in the networks or systems
with time delay. {\it Yang} [2000] has studied the effect of
time delay on the response of small-world networks.

The purpose of the present work is to model geophysical processes such as
pollutant transport, permeability model and saltwater intrusion
[{\it Holzbercher}, 1998] as  a particle percolation process
on the small-world networks with time-delay by using or modifying the
small-world theory so as to characterise the small-world networks
with time-delay by using fractal dimension and system saturation
time in the percolative media. The present model will
generally lead to a delay differential equation, whose solution is usually
very difficult to obtain if it is not impossible. Thus the numerical
simulation becomes essential. However, we will take the analytical
analysis as far as possible and compare with the results from numerical
simulations.

\section{Mathematical Model of Small-World Networks}

Considering a randomly connected network with
$n$-dimensional lattice [{\it Watts and Strogatz}, 1998; {\it Bollobas}, 1985;
{\it Watts}, 1999] and $n=1,2,3$. Overlapping on the $n$-dimensional networks
are a number of long-range shortcuts randomly connecting some vertices,
and the fraction of the long-range shortcuts or probability $p$ is relative
small $p \ll 1$. Now assuming an influence or a pollutant
particle spreads with a constant velocity $u=1$ in all directions
and a newly infected site in the other end of a shortcut will start
but with a time delay $\Delta$. Following the method
developed by {\it Newman and Watts} [1999]
and {\it Moukarzel} [1999], the total influenced volume
$V(t)$ comes from two contributions: one is the influenced volume with
$\q \int^{t}_0 t^{n-1} dt $ where $t$ is time, and the other contribution
is $2 p V(t-\z-\Delta)$ for a sphere started at time $\z$.
By using a continuum approach to the network, then $V(t)$
satisfies the following equation with time delay
\be
V(t)=\q \int^{t}_0 \z^{n-1} [1+\xi^{-n} V(t-\z-\Delta)] d\z,
\label{equ-1}
\ee
where $n=1,2,3$ and $\q$ is shape factor of a hypersphere in $n$-dimensions.
The length scale of {\it Newman-Watts} type [1999] can be defined as
\be
\xi=\k{2}{(p k n)^{1/n}},
\ee
where $k=const$ being some fixed range. By proper rescaling $t$
by $(\q u (n-1)!)^{1/n}$, that is
\be
\t=t (\q \xi^{-n} (n-1)!)^{1/n}, \s \d=\Delta (\q \xi^{-n} (n-1)!)^{1/n}.
\ee
By rewriting (\ref{equ-1}) in the rescaled form and
differentiating the equation $n$ times, we have a time-delay equation
\be
\pcd{V}{\t}=\xi^{n}+ V(\t-\d),
\ee
which is a delay differential equation, whose explicit solutions is not always
possible depending on the initial conditions. For a proper initial condition
$V(\t)=0$ for $-\d \le \t \le 0$, the solution can be explicitly written
by using the time-stepping method,
\be
V(\t)=\xi^{n} \sumj \k{(\t-j \d)^{nj}}{(nj)!}, \s n=1,2.
\label{equ-3}
\ee
Clearly, for $\t \ne 0$ and $\d \ra 0$, the our solution degenerates into
\be
S(\t)=\xi^{n} \sumjj \k{\t^{nj}}{(nj)!},
\ee
given by Moukarzel [3]. The solutions explicitly become
$V(\t)=\xi (e^{\t}-1)$ for $n=1$,
and $V(\t)=\xi^2(\cosh \t -1)$ for $n=2$.
However, when $\d \ne 0$, there is no such simple
expressions. It can be expected that the time-delay parameter $\d$ will
have a strong effect on the evolution behaviour $V(\t)$ of an influence.

\section{Pollutant Transport and Saturation Time}

By using the small-world network model, we can simulate the diffusion and
transport process
in the disordered media. Imagine a pollutant particle walk randomly
on the networks which correspond to the interconnected networks in a way
structured like the swiss-cheese model for disordered porous media, while
the longe range shortcuts corresponds to the fractures and cracks in
soil or porous rocks. One practical interest is that how long it would take
for the pollutant to reach the whole (but finite) media or the typical
saturation time of process. Obviously, the speed and efficiency of the
diffusion process depends on the permeability, and the permeability
in turn depends on the structure of porous rocks and the way of
connections. The fraction or probability $p$ of the cracks
may have strong influence on the permeability even for the very small
$p \ll 1$, which is a typical feature of small-world networks.

The evolutionary solution of $V(\t)$ with time $\t$ and time delay $\d$ gives
the volume affected by the pollutant.
Solution (\ref{equ-3}) is valid for an infinite network.
However, in reality most systems are finite. Since $V(\t)$ increases
dramatically as $\t$ increases. we can expect that the whole
system will be reached and saturated when $\t$ tends to a typical
value $\t_s$, which represents a system saturation time. For a finite
system with a size $N \sim L^n$, the characteristic saturation time
$\t_s$ can be estimated by
\be
L^n \sim \xi^n \sum_{j=1}^{\t_s/\d} \k{(\t_s-j \d)^{nj}}{(nj)!}.
\ee
In most random walk models, it often the time delay is small or $\d \ra 0$.
Now for simplicity, we take $\d=0$. For $n=1$,
we have
\be
\t_s \sim \ln (L/\xi)
\ee
or in terms of real time
$t \sim (\xi/A) \ln (L/\xi)$ where $A=(\q (n-1)!)^{1/n}$.
For $n=2$, we have
\be
\t_s \sim \cosh^{-1} (L^2/\xi^2), \s {\rm or} \s t_s \sim 4 \xi \ln (L/\xi).
\ee
In general, $\t_s$ (or $t_s$) is a function of typical length
scale $\xi$.   Substituting the express $\xi$ in terms of $p$
into the above equation, we have
\be
t_s \sim  \k{\ln (L p^{1/n})}{p^{1/n}}.
\ee
This shows that the fraction $p$ of cracks in disordered porous
rocks can influence on the saturation time significantly.
For a fixed large $L$, then $t_s \sim (\ln L)/p^{1/n}$.
Thus, the increase of $p$, say, $p=10^{-3}$ to $p=10^{-2}$ will
decrease $t_s$ by a factor of $10$ for $n=1$. This have important
implications in management of the waste disposal and pollutant control.
For example, the pollutant wastes were disposed in the area with possible
cracks even the possibility is very small, the risk increases dramatically
as the probability increases as seen from the saturation time.

This model can also apply to other processes such as the forest fire,
oil recovery and saltwater intrusion. If there is a fire start in
a forest, the sparking effect will increase the probability of the long
range shortcuts, and thus decrease the saturation time and the fire
spread more rapidly. Similarly, saltwater intrusion proceed more
quickly in the permeable zone with even small portion of cracks.

\section{Multifractals}

Since the velocity of a particle travels at a constant speed $u=1$,
we can rewrite the solution (\ref{equ-3}) in terms of the distance $r$
and $\t=r A/\xi$ with $A=(\q (n-1)!)^{1/n}$, we have
\be
V(r)=\xi^{n} \sumr \k{(Ar/\xi-j \d)^{nj}}{(nj)!}, \s n=1,2.
\label{equ-4}
\ee
Since $V(r)$ increases as $r$ increases or $V(r) \sim r^{D}$.
One can expect that fractal dimension of the small-world
networks will give a measure of the network properties as first done by
{\it Newman and Watts} [1999] which does not include the effect of time-delay.
Then the fractal dimension $D$ can be calculated by differential
formulae
\be
D=\k{d \ln V(r)}{d\ln r},
\ee
which can be calculated using (\ref{equ-4}). We now show how the time-delay
affect the fractal dimension $D$ of the small-networks.
For $\d \ra 0$ and $n=1$, we have $D \ra 1$ for $r/\xi \ll 1$
and $D \ra A  \xi^{-1}$ for $r/\xi \ra \infty$.
For $\d \ne 0$, there is no general explicit asymptotic expressions
and numerical evaluations can be easily done. The dependence of
the fractal dimension $D$ on the length scale $\xi$ and the time-delay
$\d$ suggests the multifractal features of the small-networks.

In order to compare the analytical solution with numerical simulations,
we also use numerical simulations [{\it Watts},1999].
The numerical simulations for $n=1$ and the comparison of
fractal dimension with
the analytical solution (\ref{equ-4}) is shown in Figure 1 where
the network size $N=500,000$, $r=1000$ and
$\d=0,1,5,10$. We can see that numerical results (marked with $\circ$,
$\diamond$ etc) are in good agreement with the analytical solution (solid
lines) by using (\ref{equ-4}). As expected, for $p \sim 1/\xi \ll 10^{-3}$
(or $\xi \gg r$ so that $r/\xi \ll 1$), both numerical
results and analytical fractal dimension approach to $n=1$. The time-delay
has a very strong effect on the fractal dimension $D$ of the small-world
networks. For $\d \gg 1$, $D$ is substantially reduced compared with
the one without  time-delay $\d=0$, and $D$ is quite near $n$ for most
of the region. The large time-delay actually makes the network become
larger, and the influence spread slowly in more localized areas.

\begin{figure}
\centerline{\includegraphics[width=3.5in,angle=270]{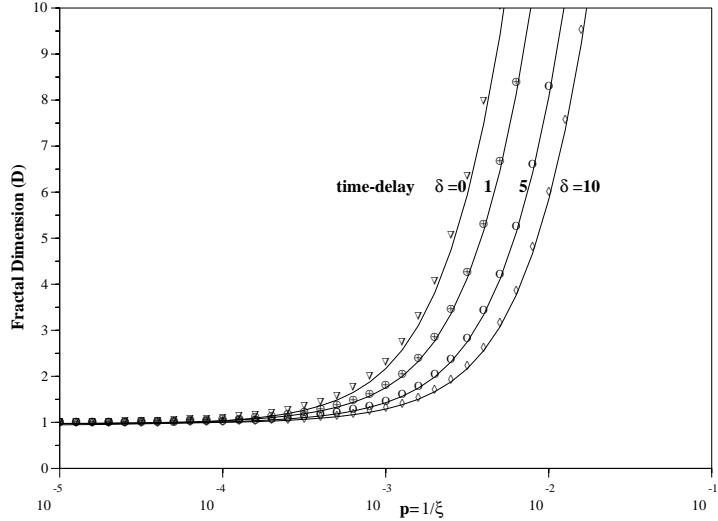}}

\caption{Fractal dimension of small-world networks with time-delay
$\d=0,1,5,10$ for a network size $N=500,000$, $r=1000$
and $n=1$. Numerical results (marked with $\circ$, $\diamond$ etc)
agree well with analytical express (solid) calculated from (11). }
\end{figure}

\section{Percolation in Porous Media}

Diffusion and transport process can be considered as
a percolation phenomenon.
In addition, permeability and conductivity can also
be modelled as a percolation
process on the interconnected networks by using Monte Carlo simulations
or fractional Brownian motion method. From the discussion in the previous
sections, we can see naturally that the small-world network model can show
newlight in the modelling of subsurface flow and contaminant transport
in disordered porous media [{\it Holzbercher}, 1998].
In small-world model, all the site or vertices
are susceptible to pollutant particles. However, in reality, there
is only a fraction of the sites can do so as in the case of site percolation.
In other words, each site is occupied with a probability or
density of $\phi$. Site percolation can also be used as the
idealized model for the distribution of oil/gas or hydrocarbon inside
porous rocks. The average concentration of fluids in the porous rocks is presented
by the occupation probability $\phi$, which happens to be the porosity
in the saturated sediments or shales. A test particle or ant in the labyrinth
move randomly on the interconnected network and various
studies [e.g., {\it Stauffer and Aharony}, 1994; {\it Yang}, 2000]
show that the percolation pattern has the fractal feature.
Now we shall discuss how $\phi$ affect the behaviour of the
small-world model.

With site occupation probability $\phi$ as an additional parameter,
the solution (\ref{equ-3}) shall be modified as
\be
V(t)=\phi \xi^{n} \sum_{j=1}^{At/(\xi \d)}
\k{(A \phi t/\xi-j \d)^{nj}}{(nj)!}, \s \phi > \phi_c,
\label{equ-7}
\ee
where $\phi_c$ is the critical probability and the above solution is
only meaningful for $\phi > \phi_c$. The critical value $\phi_c$
can be estimated numerically. However, for some type of lattice structure,
it can be caculated exactly (e.g., for 1-D lattice, $\phi_c=1$).
The introduction of a number of $p k N$ shortcuts (where $N$ is the network
size) will generally change $\phi_c$ as discovered by {\it Newman and Watts}
[1999] by solving the 1-D case
\be
p=\k{(1-\phi_c)^k}{2 k \phi_c}.
\ee
For 2-D or higher dimension case, there is no general explicit solution for
$\phi_c$ and numerical approach is usually used. For a lattice
in $n$-dimensional percolation structure,
formation of one very large cluster gives
\be
\phi_c^2 p k N=\k{N}{2} \sum_{s} g_{ks} \phi_c^k (1-\phi_c)^{s}],
\ee
or
\be
p=\k{1}{2 k} \sum_{s} g_{ks} \phi_c^{k-2} (1-\phi_c)^{s}],
\ee
where the summation is over $s$ which is the cluster perimeter
[{\it Stauffer and Aharony}, 1994].
The coefficient $g_{ks}$ depends on the network or lattice structure.
For example, the triplet on a square lattice corresponds to $k=3$ and
$s=8$ with $g_{ks}=2$ or $s=7$ with $g_{ks}=4$. A non-zero $p$ usually
gives a lower value of $\phi_c$ than that for $p=0$, which means that
cracks or higher permeability connection will reduce the critical
site occupation probability and thus increase the efficiency of the
transport and diffusion process in disordered porous media.

\section{Discussion}

A modified small-world network model has been presented here to
characterise the effect of time-delay on small-world networks.
Numerical simulations and analytical analysis for networks
with time-delay show that the time-delay parameter $\d$ has
a very strong effect on
the fractal dimension and other properties such as the speed and saturation
time of the delay networks. A small-world network can become
larger if sufficient time-delay in the system response in introduced.

On the other hand, in order to make a large world network transfer into
a small-world or to increase the transport and diffusion efficiency of
the network,  a slight higher probability $p$ of long-range
random shortcuts are necessary compared with the one without time-delay
because this essentially reduce the characteristic length scale $\xi=O(1/p)$
quite significantly. Cracks and other long-range connections can have a
strong influence on permeability, conductivity and percolation
in disordered porous media.
Further studies of the dynamics of small-world
networks are obviously necessary and the present work is just one
of such attempts to the understanding of dynamics of networks and
application in geophysical modelling.

\section{References}

\begin{description}
\item M Barthelemy and L A N Amaral,
      Small-world networks: Evidence for a crossover picture
      {\it Phys. Rev. Lett.},  {\bf 82}, 3180-3183 (1999).

\item B Bollobas, {\it Random graphs}, Academic Press, New York, 1985.

\item M Boots and A Sasaki,
      Small worlds and the evolution of virulence:
      infection occurs locally and at a distance, {\it
      P Roy Soc Lond}, B {\bf 266},1933-1938(1999).

\item E O Holzbercher, {\it Modeling Density-Driven Flow in Porous Media: Principles, Numerics, Software},Springer-Verlag Berlin, 1998.

\item M E J Newman and D J Watts, Scaling and percolation in the small-world
network model, {\it Phys. Rev.}, E {\bf 60}, 7332-7342 (1999).

\item M E J Newman, C Morre and D J Watts, Mean-field solution of the
small-world network model, {\it Phys. Rev. Lett.}, {\bf 84}, 3201-3204 (2000).

\item C F Moukarzel, Spreading and shortest paths in systems with sparse
long-range connections, {\it Phys. Rev.}, E {\bf 60}, R6263-6266 (1999).

\item S A Pandit and R E Amritkar,
      Characterization and control of small-world networks
      {\it Phys. Rev.}, E {\bf 60}, R1119-1122 (1999).

\item D Stauffer and A Aharony, {\it Introduction to percolation theory},
2nd ed., Taylor and Francis, 1994.

\item D J Watts and S H Strogatz, Collective dynamics of small-world networks,
{\it Nature} (London), {\bf 393}, 440-442 (1998).

\item D J Watts, {\it Small worlds: The dynamics of networks between order and randomness}, Princeton Univ. Press, 1999.

\item X S Yang, Small-world networks: Effect of time-delay, {\it Europhys.
Lett.},  2000 (in press).

\end{description}

\end{document}